\documentstyle[11pt,twoside,colloqOHP2010,epsfig]{article}
\begin{document}
\title{A search for transit timing variation}
\author{G.Maciejewski$^1$, R.Neuh\"auser$^1$, St.Raetz$^1$, R.Errmann$^1$, U.Kramm$^2$ \& T.O.B.Schmidt$^1$}
\affil{$^1$ Astrophysikalisches Institut und Universit\"ats-Sternwarte, Schillerg\"asschen 2-3, D-07745 Jena, Germany [gm@astro.uni-jena.de]} 
\affil{$^2$ Institut f\"ur Physik, Univ. Rostock, D-18051 Rostock, Germany} 
\begin{abstract}
Photometric follow-ups of transiting exoplanets (TEPs) may lead to discoveries of additional, less massive bodies in extrasolar systems. This is possible by detecting and then analysing variations in transit timing of transiting exoplanets. In 2009 we launched an international observing campaign, the aim of which is to detect and characterise signals of transit timing variation (TTV) in selected TEPs. The programme is realised by collecting data from 0.6–-2.2-m telescopes spread worldwide at different longitudes. We present our observing strategy and summarise first results for WASP-3b with evidence for a 15 Earth-mass perturber in an outer 2:1 orbital resonance.
\end{abstract}
\section{Introduction}
Timing of transiting exoplanets is expected to provide discoveries of additional, even very low mass bodies in extrasolar systems (e.g. Holman \& Murray 2005). In a single-planet system, a transiting planet orbits its host star on a Keplerian orbit. If there is another planet in the system, it interacts gravitationally with the transiting planet what generates deviations from the strictly Keplerian case. These perturbations result in a quasi-periodic signal in an observed-minus-calculated (O--C) diagram of the transiting planet. The TTV method is sensitive to small perturbing masses in orbits near to the low-order mean-motion resonances. Deriving the orbital elements and mass of the perturber from the TTV signal is a difficult inverse problem. Different configurations may generate similar TTV characteristics with an identical dominant periodicity. For a given transiting planet, the TTV signal depends on perturber's mass and its orbital elements. Exploring such a multi-dimensional space of parameters is not a trivial task and systematic observations are crucial to finding a correct solution.
\section{Strategy}
Data from telescopes located at various longitudes are needed to observe most target's transits which are available in an observing season. Thus, we organise multi-site campaigns which engage telescopes in Europe, Asia and America. Our strategy may be summarised as follows:
\begin{enumerate}
	\item Choosing a target sample: the most promising targets are being selected by reanalysis of available literature data. We observe only 2--3 targets in a given season to maximise the efficiency of the project.
\item Reconnaissance campaign: an attempt to detect TTV with telescopes smaller than 1 m. If a TTV signal is detected, a provisional hypothesis is put forward. Models, which assume the existence of a perturber in a system and reproduce observed variation of timing residuals, are identified by three-body simulations.
\item Follow-up campaign: a photometric follow-up is run with big telescopes ($>$1 m) to verify and characterise TTV signal. In some cases a spectroscopic follow-up is organised to get more precise radial velocity measurements. If a claimed TTV signal is confirmed, a final hypothesis is elaborated, based on both high-precision photometric and spectroscopic data.
\end{enumerate}

\section{First results}

\begin{figure}[t]
\begin{center}
\epsfig{width=7.0cm,file=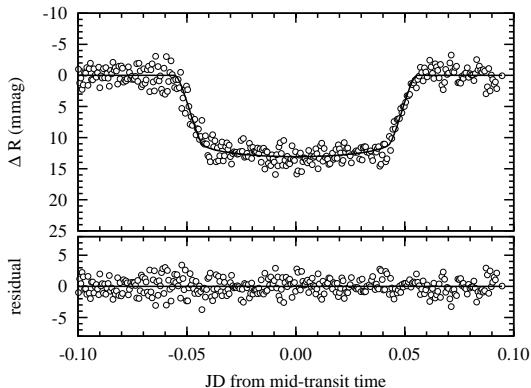}
\caption{An R-band light curve of the WASP-3b transit observed on 23 July 2010 with the 2.2-m telescope at Calar Alto (Spain). Timing error is 25 s and photometric precision is 0.9 mmag despite non-photometric conditions. The Exoplanet Transit Database (Poddan\'y, Br\'at, \& Pejcha 2010) was used to obtain the best-fitting model.}
\end{center}
\end{figure}

In 2009 and 2010, we observed 6 transits of exoplanet WASP-3b during a pilot campaign. The team used the 90-cm telescope of the University Observatory Jena (Germany) and the 60-cm telescope of the Rohzen National Astronomical Observatory (Bulgaria). We noted that the transit timing cannot be explained by a constant period but by a periodic variation in the O--C diagram. Simplified models assuming the existence of a perturbing planet in the system and reproducing the observed variations of timing residuals were identified by three-body simulations. We found that the configuration with the hypothetical second planet of mass of about 15 Earth masses, located close to the outer 2:1 mean motion resonance, is the most likely scenario reproducing observed transit timing (Maciejewski \etal~2010). 

In summer 2010, the dedicated follow-up campaign has begun. New radial-velocity measurements have been gathered with the Hobby-Eberly Telescope (USA). High-precision light curves have been obtained with 2-m class telescopes in Europe, North America and Asia to determine timing with errors smaller than 30 s. Fig.~1 shows the exemplary light curve of a complete transit of WASP-3b, obtained during the follow-up programme. Results of the ongoing campaign are expected to confirm the claimed TTV signal and put tighter constraints on parameters of the perturbing body. A long-time baseline of observations will allow to study possible resonant oscillations of the planetary orbits.

\acknowledgments{GM acknowledges support from the EU in the FP6 MC ToK project MTKD-CT-2006-042514. GM and RN acknowledge support from the DAAD PPP--MNiSW project 50724260--2010/20011. SR, UK and TS acknowledge support from DFG in programs NE 515/32-1, SPP 1385 and NE 515/30-1, respectively.}


\begin{references}
\reference Holman, M.J., \& Murray, N.W. 2005, Science, 307, 1288
\reference Maciejewski, G., Dimitrov, D., Neuh\"auser, R., \etal~2010, MNRAS, 407, 2625
\reference Poddan\'y, S., Br\'at, L., \& Pejcha, O. 2010, New Astron., 15, 297
\end{references}
\end{document}